\documentclass[preprints,accept,oneauthor,pdftex,10pt,a4paper]{Definitions/mdpi}

\firstpage{1}
\articlenumber{x}
\doinum{10.3390/------}
\pubvolume{xx}
\pubyear{2018}
\copyrightyear{2018}
\history{Received: date; Accepted: date; Published: date}

\pdfoutput=1

\usepackage{amssymb}

\Title{NA61/SHINE experiment - programme beyond 2020\footnote{Lecture given at the Colloqium on Nonequilibrium Phenomena in Strongly Correlated Systems, Dubna, 18 - 19 April, 2018}}

\Author{Ludwik Turko\\ for the NA61/SHINE Collaboration}

\AuthorNames{Ludwik Turko}

\address[1]{Institute of Theoretical Physics, University of Wroclaw, pl. M.~Borna 9, 50-205 Wroclaw, Poland; ludwik.turko@ift.uni.wroc.pl}

\corres{Correspondence: ludwik.turko@ift.uni.wroc.pl}

\abstract{
The fixed-target NA61/SHINE experiment (SPS CERN) looks for the critical point (CR) of strongly interacting matter and the properties of the onset of deconfinement. It is a scan of measurements of particle spectra and fluctuations in proton-proton, proton-nucleus and nucleus-nucleus interactions as a function of collision energy and system size, corresponding to a two dimensional phase diagram (T-$\mu_B$). New measurements and their objectives, related to the third stage of the experiment after 2020 are presented and discussed here.
}


\begin{document}
\section{Introduction}
\begin{figure}[h!]
\centering
\includegraphics[width=0.8\textwidth]{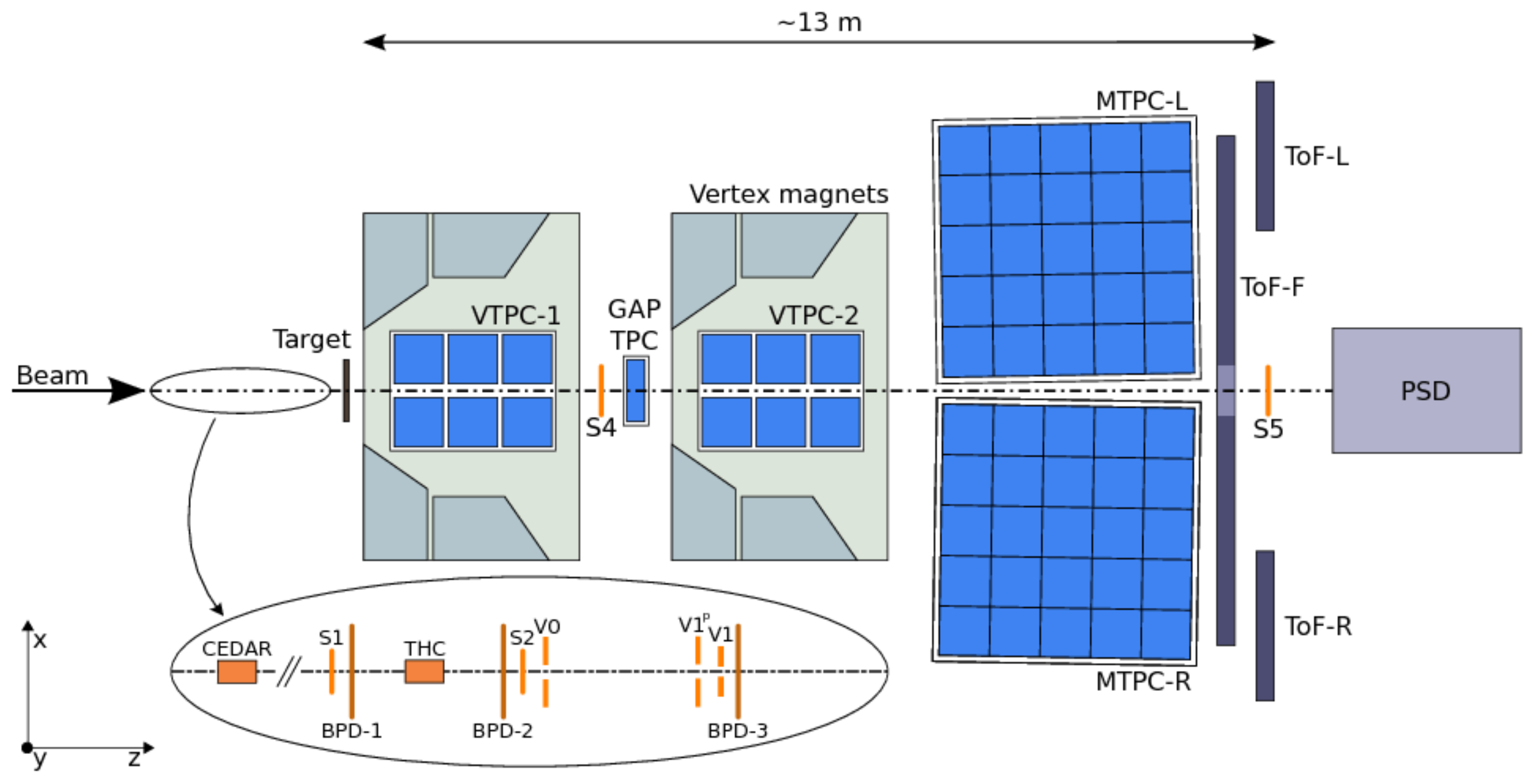}
\caption{The present NA61 / SHINE detector consists of a large acceptance hadron spectrometer followed by a set of six Time Projection Chambers as well as Time-of-Flight detectors. The high resolution forward calorimeter, the Projectile Spectator Detector, measures energy flow around the beam direction. For hadron-nucleus interactions, the collision volume is determined by counting low momentum particles emitted from the nuclear target with the LMPD detector (a small TPC) surrounding the target. An array of beam detectors identifies beam particles, secondary hadrons and nuclei as well as primary nuclei, and measures precisely their trajectories}\label{NA61}
\end{figure}

The NA61/SHINE, \textbf{S}uper Proton Synchrotron (SPS) \textbf{H}eavy \textbf{I}on and \textbf{N}eutrino \textbf{E}xperiment) is the continuation and extension of the NA49 \cite{Antoniou:2006_034,  Abgrall:2008_018} measurements of hadron and nuclear fragment production properties in fixed-target reactions induced by hadron and ion beams. It has used a similar experimental fixed-target setup as NA49 (Figure \ref{NA61})  but with an  extended research programme. Beyond an enhanced strong interactions programme there are the measurements of hadron-production for neutrino and cosmic ray experiments realized \cite{Turko:2018kvt}. NA61/SHINE is the collaboration with about 150 physicists, 30 institutions and 14 countries being involved.

\begin{figure}[h!]
\centering
\includegraphics[width=0.4\textwidth]{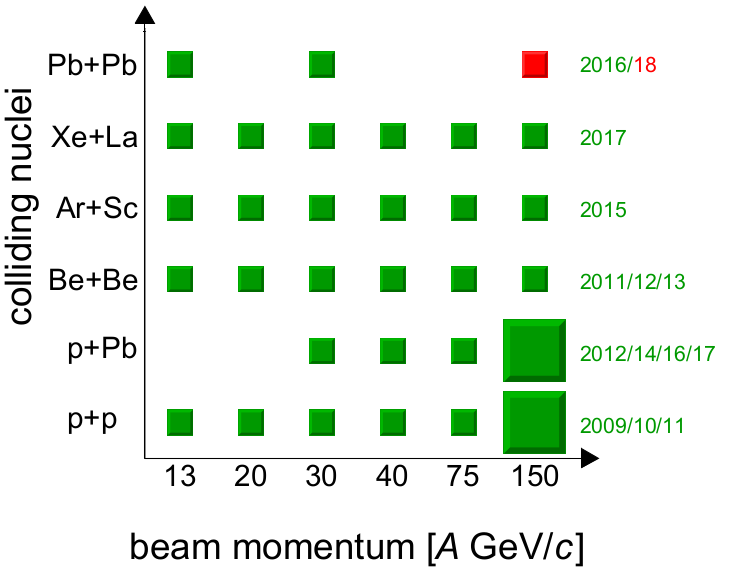}
\caption{For the programme on strong interactions NA61/SHINE scans in the system size and beam momentum. In the plot the recorded data are indicated in green, the approved future data taking in red}\label{beams}
\end{figure}

The strong interaction programme of NA61/SHINE is devoted to study the onset of deconfinement and search for the CR of hadronic matter, related to the phase transition between hadron gas (HG) and quark-gluon plasma (QGP). The NA49 experiment studied  hadron production in Pb+Pb interactions while the NA61/SHINE collects data varying collision energy (13A-158A GeV) and size of the colliding systems. This is equivalent to the two dimensional scan of the hadronic phase diagram in the $(T, \mu_B)$ plane, as depicted in Figure \ref{beams}. The research programme was initiated in 2009 with the p+p collisions used later on as reference measurements for heavy-ion collisions.

Hadron-production measurements for neutrino experiments are just reference measurements of p+C interactions for the T2K experiment, since they are necessary for computing initial neutrino fluxes at J-PARC. These measurements have been extended to the production of charged pions and kaons in interactions with thin carbon targets, and replicas of the T2K target, to test accelerator neutrino beams accelerator neutrino beams \cite{Abgrall:2011ae}. Data taking began in 2007.

Collected p+C data allows also for better understanding of nuclear cascades in the cosmic air showers - necessary in the Pierre Auger and KASCADE experiments \cite{Auger_2004,Kascade_2003}. These are reference measurements of p+C, p+p, $\pi$+C, and K+C interactions for cosmic ray physics. The cosmic ray collisions with the Earth's atmosphere  produce air shower secondary radiation. Some of particles produced in such collisions subsequently decay into muons, which are able to reach the surface of the Earth.  Cosmic ray induced muon production would allow to reproduce primary cosmic ray composition if related hadronic interaction are known \cite{Morison_2008}.

\begin{figure}[h!]
  \centering
  \includegraphics[width=0.5\textwidth]{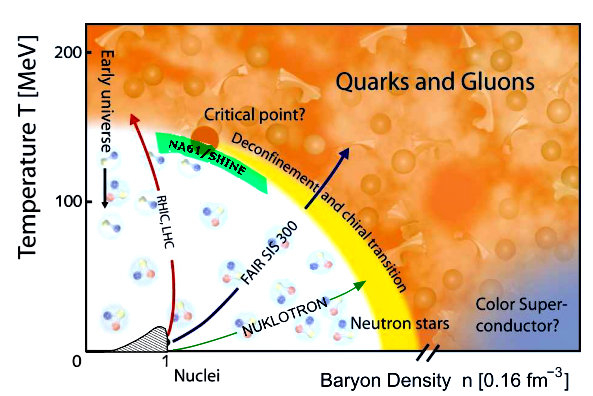}
  \caption{Phase structure of hadronic matter covered by NA61/SHINE (green) as compared to present and future heavy ion experiments}
\label{Phases}
\end{figure}

As seen in Figure \ref{Phases},  phase  structure  of  hadronic  matter  is involved. Progresses in theoretical understanding of the subject and collecting more
experimental data will allow delving into the subject. While the highest energies achieved at LHC and RHIC colliders provide data related to the crossover HG/QGP regions, the SPS fixed-target NA61/SHINE experiment is particularly suited to explore the phase transition line HG/QGP with the CR included.

\subsection{Results of initial NA61/SHINE research}

Production properties of light and medium mass hadrons, in particular pions and kaons, have been measured, according to the NA61/SHINE proposal \cite{Antoniou:2006_034}. The Be+Be results are close to p+p independently of collision energy. Moreover, the data show a jump between light (p+p, Be+Be) and intermediate, heavy (Ar+Sc, Pb+Pb) systems \cite{Aduszkiewicz:2017_038}.  The observed rapid change of hadron production properties that start when moving from Be+Be to Ar+Sc collisions can be interpreted as the beginning of creation of large clusters of strongly interacting matter - the onset of fireball. One notes that non-equilibrium clusters produced in p+p and  Be+Be collisions seem to have similar
properties at all beam momenta studied here.

 The $K^+/\pi^+$ ratio in p+p interactions is below the predictions of statistical models.
However, the ratio in central Pb+Pb collisions is close to statistical model predictions for large volume system \cite{Becattini:2006}.

In p+p interactions, and thus also in Be+Be collisions, multiplicity fluctuations are
larger than predicted by statistical models. However, they are close to statistical
model predictions for large volume systems in central Ar+Sc and Pb+Pb collisions \cite{Begun:2007}.

\begin{figure}[!h]
\includegraphics[width=0.4\linewidth]{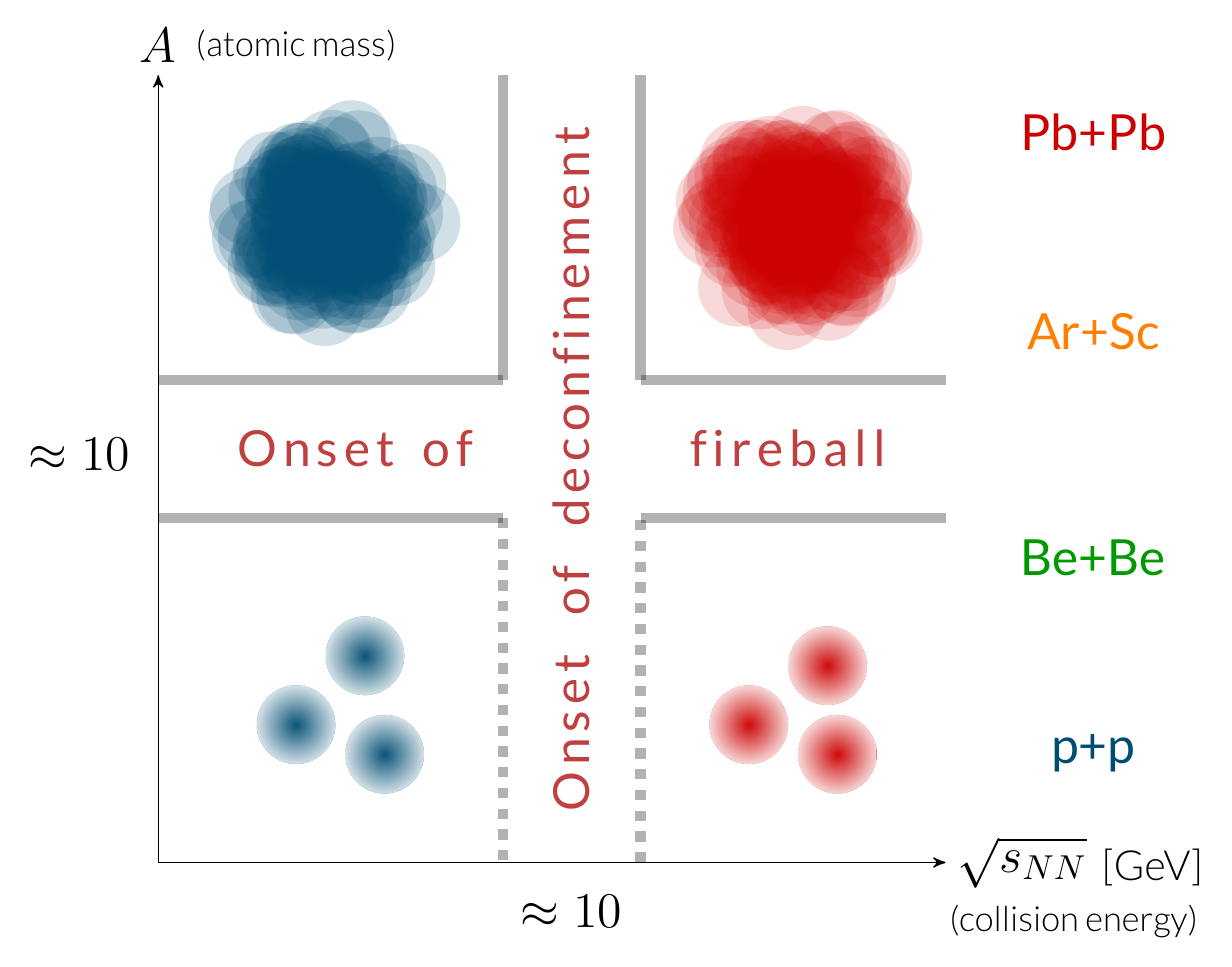}
\caption{The onset of deconfinement and the onset of fireball. The onset of deconfinement is well established in central Pb+Pb (Au+Au) collisions, its presence in collisions of low mass nuclei, in particular, inelastic p+p interactions is questionable.}
\label{cluster}
\end{figure}

The two-dimensional scan conducted by NA61/SHINE by varying collision energy and nuclear mass number of colliding nuclei indicates four domains of hadron production separated by two thresholds: the onset of deconfinement and the onset of fireball. The sketch presented in Fig.\ref{cluster} illustrates this conclusion.

Total production cross sections and total inelastic cross sections for reactions $\pi^+$+C,Al and $K^+$+C,Al  at 60~GeV/c and $\pi^+$+C,Al at 31~GeV/c were measured. These measurements are a key ingredient for neutrino flux prediction from the reinteractions of secondary hadrons in current and future accelerator-based long-baseline neutrino experiments \cite{Aduszkiewicz:2018uts}.

A more detailed exploration of QCD phase diagram would need both new experimental data with extended detection capabilities and improved theoretical models \cite{Caines:2017vvs}.

\section{New measurements requested}

The third stage of the experiment, starting after the Long Shutdown 2 (LS-2) of the CERN accelerator system,   would include

\begin{itemize}
  \item measurements of charm hadron production in Pb+Pb collisions for heavy ion physics;

 \item measurements of nuclear fragmentation cross section for cosmic ray physics;

\item measurements of hadron production in hadron-nucleus interactions for neutrino
physics.

\end{itemize}

The proposed  measurements and analysis  are requested by heavy ion, cosmic ray and neutrino communities.

The objective of \textbf{charm hadron production measurements} in Pb+Pb collisions is to obtain the first data on mean number of $\bar{c}c$ pairs produced in the full phase space in heavy ion collisions. Moreover, further new results on the collision energy and system size dependence will be provided. This will help to answer the questions about  the mechanism of open charm production, about the relation between the onset of deconfinement and open charm production, and about the behaviour of $J/\psi $ in quark-gluon plasma.

The objective of \textbf{nuclear fragmentation cross section measurements} is to provide high-precision data needed for the interpretation of results from current-generation cosmic ray experiments.  The proposed measurements are of crucial importance to extract the characteristics of the diffuse propagation of cosmic rays in the Galaxy.

The objectives of \textbf{new hadron production measurements for neutrino physics} are to improve further the precision of hadron production measurements for the currently used T2K replica target, to perform measurements for a new target material, both for T2K-II and Hyper-Kamiokande experiments, to study the possibility of measurements at low incoming beam momenta (below 12 GeV/c) relevant for improved predictions of both atmospheric and accelerator neutrino fluxes.

\begin{figure}[!h]
\includegraphics[width=0.5\linewidth]{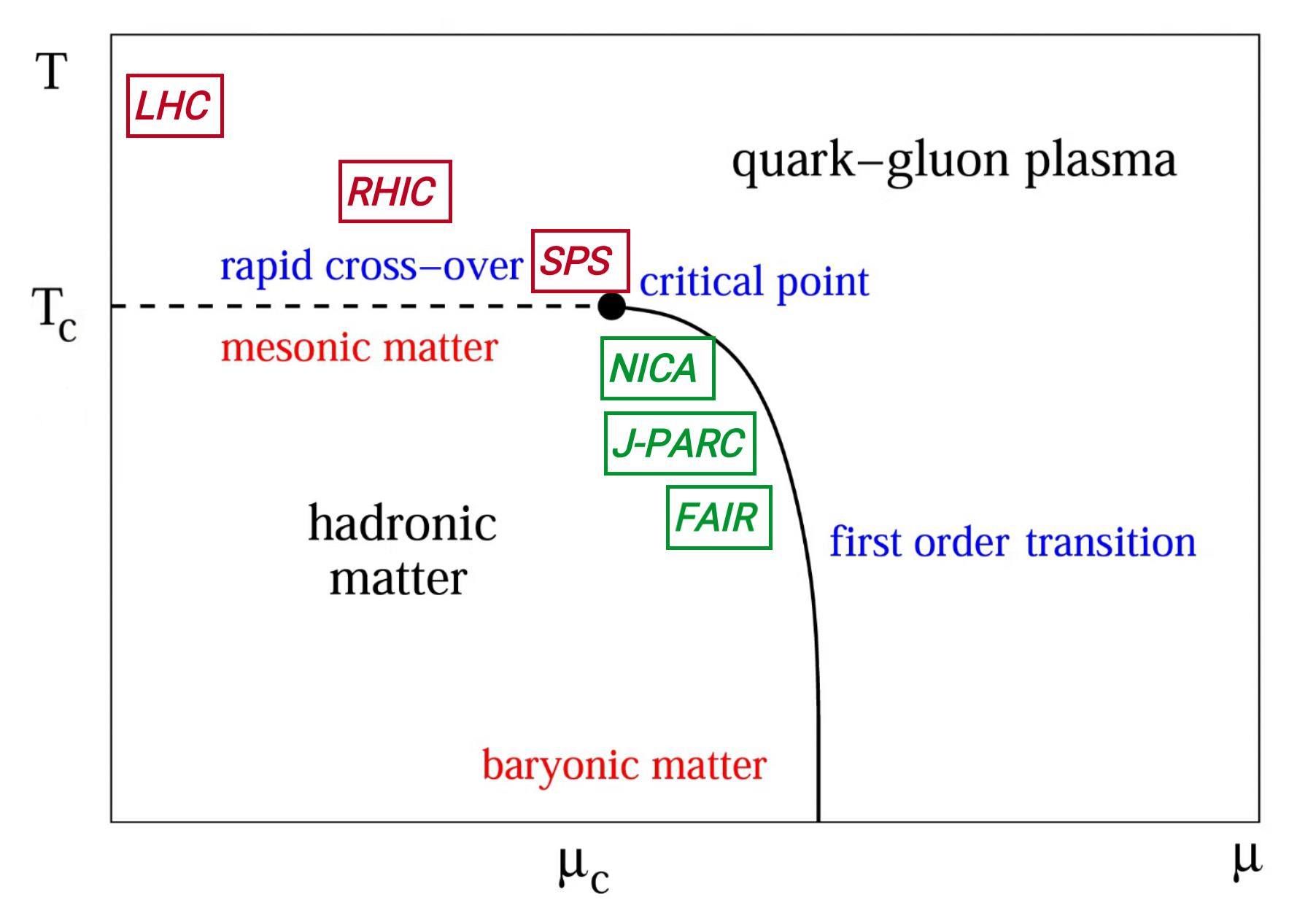}
\caption{ \small{Recent (red) and future (green) heavy ion facilities in the phase diagram
of strongly interacting matter.}}
\label{Landscape}
\end{figure}

NA61/SHINE is the only experiment which will conduct such measurements in the near future. Together with other HIC experiments creates a full-tone physical picture of QCD in dense medium.   Especially, concerning  strong  interaction heavy-ion program the NA61/SHINE has unique capabilities  in comparison with the other experiments (see Fig. \ref{Phases}):

Limitations of other experiments are related to: (i) limited acceptance, (ii) measurement of open charm not considered in the current program, (iii) or very low cross-section at SIS-100:

Concerning other experiments capabilities shown at Fig. \ref{Landscape}:

\begin{itemize}

  \item LHC and RHIC measurements of open charm at high energies are performed in a limited acceptance due to the collider kinematics and to  the detector geometry. The NA61/SHINE measurement will not be subject to these limitations. \cite{Meninno:2017, Hou:2016, Simko:2017, Nagashima:2017};

  \item RHIC BES collider ( $\sqrt{s_{NN}}=7.7-39$ GeV ): measurement not considered in the current program \cite{Odyniec:2013, Yang:2017};

   \item RHIC BES fixed-target ( $\sqrt{s_{NN}}= 3- 7.7$ GeV ): measurement not considered in the current program \cite{Meehan:2016};

  \item  NICA ( $\sqrt{s_{NN}} < 11$ GeV): measurements during stage 2 (after 2023) are under consideration \cite{Kekelidze:2017tgp};

    \item J-PARC-HI ( $\sqrt{s_{NN}}\lesssim 6$ GeV): under consideration, may be possible after 2025 \cite{Sako:2016edz};

     \item FAIR SIS-100 ( $\sqrt{s_{NN}}\lesssim 5$): not possible due to the very low cross-section at SIS-100, charm measurements are planned with SIS-300 ($\sqrt{s_{NN}}\lesssim 7$ GeV), but not of the start version (timeline is unclear). 
\end{itemize}

The beam momentum range provided to NA61/SHINE by the SPS and the H2 beam line is highly important for the heavy ion, neutrino and cosmic ray communities. It covers:
\begin{itemize}
   \item  energies at which the transition from confined hadrons to quark gluon plasma in heavy ion collisions takes place - the onset of deconfinement \cite{Gazdzicki:2010iv};
    \item  proton beams of momenta used to produce neutrino beams at J-PARC, Japan and Fermilab, US \cite{Abgrall:2018nu};
  \item light nuclei at momenta $>$ 10 A GeV/c, important for the understanding of the propagation of cosmic rays in the Galaxy \cite{Genolini:2018ekk}.
\end{itemize}

\subsection{Specific research goals}

The NA61/SHINE charm program addresses questions of the validity and the limits of statistical and dynamical models of high energy collisions in the new domain of quark mass, $m_c \approx 1300\ \mbox{MeV} \gg T_C \approx 150\ \mbox{MeV}$  \cite{Snoch:2018nnj}. To answer these questions, knowledge is needed on the mean number of charm–anti-charm quark pairs $\langle c\bar{c}\rangle$  produced in the full phase space of heavy ion collisions.

\begin{figure}[!h]
\includegraphics[width=0.9\linewidth]{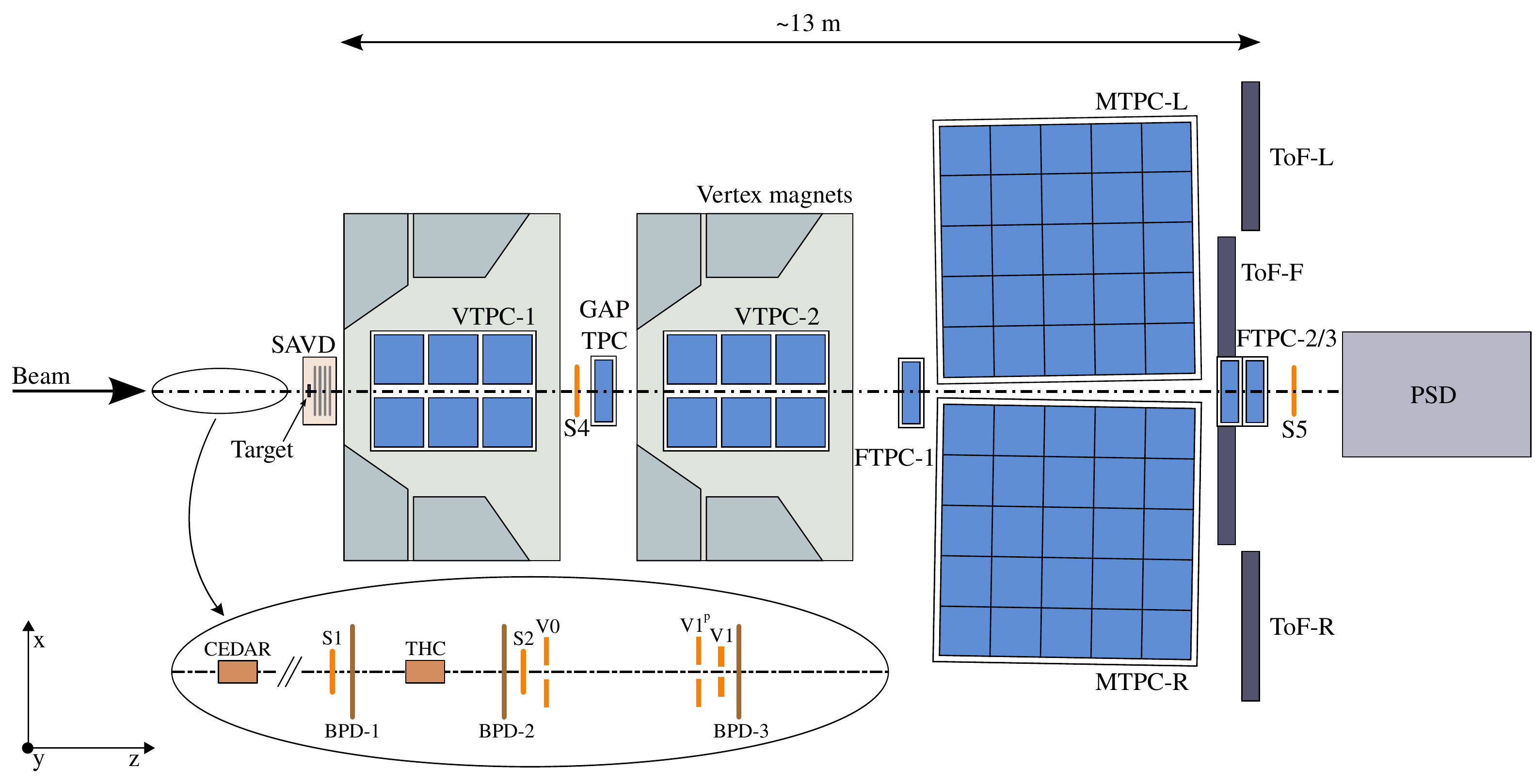}
\caption{\small{Present NA61/SHINE setup with the SAVD included}}
\label{Setup}
\end{figure}

Such data do not exist yet and NA61/SHINE aims to provide them within the coming years. The related preparations have started already. In 2015 and 2016, a Small Acceptance Vertex Detector (SAVD) was constructed and first measurements of open charm production started in 2016 - Fig. \ref{Setup}. Vertex resolution has appeared precise enough (30 $\mu$m) to distinguish $D^0$ decay. That was possible due to the fixed target experiment specific property, where the Lorenz factor $\beta\gamma\approx 10$ makes short-living $D^0$ an observable particle even in such a small acceptance vertex detector.

Successful performance of the SAVD in 2016 led to the decision to also use it during the Xe+La data taking in 2017. About $5*10^6$ events of central Xe+La collisions at 150A GeV/c
were collected. The Xe+La data are currently under analysis and are expected to lead to physics results in the coming months. One expects to reconstruct several hundred of $D^0$
and $\bar{D^0}$  decays. Beyond this about 4000 $D^0$ and $\bar{D^0}$  decays can be expected to be reconstructed from Pb+Pb data taking in 2018. Further data taking on Pb+Pb collisions and reconstruction of decays of various open charm mesons are planned by NA61/SHINE for the years $2022 - 2024$. This would be combined with the required detector upgrades.

Another domain of NA61/SHINE activity will be  to measure fragmentation cross sections relevant for the production of Li, Be, B, C and N nuclei. These elements are of particular importance for the physics of cosmic rays in the Galaxy. The NA61/SHINE facility has already successfully taken data with light ion beams \cite{Kaptur:2015} and can be used with practically no modifications to perform the needed cross section measurements at isotope level.  The ability to separate different isotopes from fragmentation interactions for a given charge was validated with simulations \cite{Aduszkiewicz:2017_035}.

\acknowledgments{The author acknowledges support by the Polish National Science Center under contract No. UMO-2014/15/B/ST2/03752,  the COST Actions CA15213 (THOR), CA16117 (ChETEC) and  CA16214 (PHAROS) for supporting their networking activities, and  the Bogolubov-Infeld Program for supporting author's stay at JINR Dubna.}
\abbreviations{The following abbreviations are used in this manuscript:\\
\noindent
\begin{tabular}{@{}ll}
CR & critical point\\
HG & hadron gas\\
HIC & heavy-ion collision\\
LS & long shutdown\\
QCD & quantum chromodynamics\\
QGP & quark-gluon plasma\\
SAVD & small acceptance vertex detector\\
\end{tabular}}

\reftitle{References}

\end{document}